\documentclass[aps,preprint,nofootinbib,floatfix]{revtex4}
\usepackage{graphicx,color}
\usepackage[latin1]{inputenc}
\usepackage{amsmath,amssymb}
\usepackage{hyperref}
\usepackage{epstopdf}
\usepackage{slashed}
\usepackage{soul}
\usepackage{amssymb}
\usepackage[normalem]{ulem}
\usepackage[caption=false]{subfig}

\newcommand{\lsim}{\mathrel{\mathop{\kern 0pt \rlap
  {\raise.2ex\hbox{$<$}}}
  \lower.9ex\hbox{\kern-.190em $\sim$}}}
\newcommand{\gsim}{\mathrel{\mathop{\kern 0pt \rlap
  {\raise.2ex\hbox{$>$}}}
  \lower.9ex\hbox{\kern-.190em $\sim$}}}

\newcommand{\gtwo}{\ensuremath{\delta a_\mu}}
\newcommand{\sigsip}{\ensuremath{\sigma^{\rm{SI}}_p}}

\newcommand{\gev}{\ensuremath{\,\mathrm{GeV}}}
\newcommand{\tev}{\ensuremath{\,\mathrm{TeV}}}

\def  \bcen   {\begin{center}}
\def  \ecen   {\end{center}}
\def  \beq    {\begin{equation}}
\def  \eeq    {\end{equation}}
\def  \beqa   {\begin{eqnarray}}
\def  \eeqa   {\end{eqnarray}}

\def\bea{\begin{eqnarray}}
\def\eea{\end{eqnarray}}

\begin{document}

\title{NMSSM neutralino dark matter for CDF II $W$-boson mass and muon $g-2$ and the promising prospect of direct detection}
\author{Tian-Peng Tang$^{a,b}$}
\author{Murat Abdughani$^{a,c}$}
\author{Lei Feng$^{a,b,d}$}
\author{Yue-Lin Sming Tsai$^a$}
\author{\\Jian Wu$^{a,b}$}
\author{Yi-Zhong Fan$^{a,b}$}

\affiliation{$^a$Key Laboratory of Dark Matter and Space Astronomy, 
   Purple Mountain Observatory, Chinese Academy of Sciences, Nanjing 210033, China}
\affiliation{$^b$School of Astronomy and Space Science, University of Science and Technology of China, Hefei, Anhui 230026, China}
\affiliation{$^c$School of Physical Science and Technology, Xinjiang University,
Urumqi, Xinjiang 830046, China}
\affiliation{$^d$Joint Center for Particle, Nuclear Physics and Cosmology,  
Nanjing University -- Purple Mountain Observatory,  Nanjing  210093, China}

\date{\today}

\begin{abstract}
Two experiments from the Fermilab, E989 and CDF II, have reported two anomalies for muon $g-2$ and $W$-boson mass that may indicate the new physics at the low energy scale.
Here we examine the possibility of a common origin of these two anomalies in the Next-to-Minimal Supersymmetric Standard Model. Considering various experimental and astrophysical constraints such as the Higgs mass, collider data, flavor physics, dark matter relic density, and direct detection experiments, we find that lighter electroweakinos and sleptons can generate sufficient contributions to muon $g-2$ and $m_W$. Moreover, the corresponding bino-like neutralino dark matter mass is in the $\sim 180-280$ GeV range.
Interestingly, the favored DM mass region can soon be entirely probed by ongoing direct detection experiments like PandaX-4T, XENONnT, LUX-ZEPLIN, and DARWIN.\\



\end{abstract}

\maketitle

\section{Introduction \label{sec:intro}}

The discovery of the Higgs boson completes the Standard Model (SM), and the next important task for particle physics is to search for new physics beyond the Standard Model (BSM). Although the Large Hadron Collider (LHC) offers the most direct way to find new particles, there is still no clear signal.
On the other hand, electroweak precision observables, 
such as the muon anomalous magnetic moment $a_\mu=(g-2)_\mu/2$ and the $W$-boson mass $m_W$, 
are susceptible to the loop contributions of the new particles. 
Once the deviations between SM predictions and precision measurements are confirmed, one can explore the BSM indirectly.


The experiment E989 at Fermilab has recently reported a muon anomalous magnetic moment with a relative precision of 368 parts per billion (ppb). 
The combination with the Brookhaven National Lab (BNL)~\cite{Tanabashi:2018oca} data yields a $\delta a_\mu= (2.51 \pm 0.59) \times 10^{-9}$, in tension with the SM prediction at the statistical significance of $4.2\sigma$~\cite{Aoyama:2020ynm,2104.03281}. 
Very recently, the CDF collaboration has reported their precise measurement of $W$-boson mass $m_{W, {\rm CDF}}= 80.4335 \pm 0.0094 {\rm GeV}$~\cite{CDF:2022hxs} 
using approximately 4 million $W$-boson candidates from CDF II detector data corresponding to an integrated luminosity of $\mathcal{L}= 8.8$~fb$^{-1}$ collected in $p\bar{p}$ collisions at a $\sqrt{s}$ = 1.96 TeV center-of-mass energy.
The deviation from the SM prediction $m_{W, {\rm SM}}=80.361\pm 0.006\gev$~\cite{ParticleDataGroup:2020ssz} 
is at a confidence level of $\sim 7\sigma$~\cite{CDF:2022hxs}. 
Soon after this anomaly was reported, several different new physics models have been considered to interpret the CDF II $W$-boson mass~\cite{Fan:2022dck,Zhu:2022tpr,Yuan:2022cpw,Yang:2022gvz, Lu:2022bgw,Athron:2022qpo, deBlas:2022hdk, Strumia:2022qkt}.

The Minimal Supersymmetric Standard Model (MSSM) can naturally accommodate 
the Dark Matter (DM) candidate, the hierarchy problem, and the gauge coupling unification. 
The superpotential of the MSSM contains a term bilinear in the two Higgs doublets (i.e., $\sim\mu H_u H_d$), 
but the $\mu$ parameter is so tiny compared with the Supersymmetry (SUSY) breaking scale (i.e., the so-called $\mu$-problem).
However, this puzzling problem can be solved naturally in the Next-to-Minimal Supersymmetric Standard Model (NMSSM), an extension of the MSSM with an additional gauge singlet Higgs ($S$), which generates the $\mu$-term dynamically 
through the singlet Higgs vacuum expectation value. 
In other words, the corresponding term in the superpotential is replaced by a coupling mode (i.e., $\sim \lambda S H_u H_d$). 
Compared with the MSSM, the additional particle content in the NMSSM is not only one scalar and one pseudoscalar Higgses, 
but also their superpartner Singlino. 
Hence, besides solving the $\mu$-problem, it is phenomenologically richer 
in both the Higgs sectors (three scalars, two pseudoscalars, and one charged Higgses) 
and the neutralino sector (five neutralinos).  


Considering $R$-parity conserved, the lightest neutralino can serve as a DM candidate, but its interactions with the SM are basically fixed by the Planck relic density measurement~\cite{Planck:2018vyg}.
In addition, the DM indirect detection experiments also reported some intriguing excesses, such as the Fermi Large Area Telescope (Fermi-LAT) Galactic Center $\gamma$-ray excess (GCE)~\cite{Hooper:2010mq,Zhou:2014lva,Calore:2014xka,Daylan:2014rsa} and the Alpha Magnetic Spectrometer (AMS-02) experiment antiproton excess~\cite{1610.03840,1610.03071}. 
Within the framework of the NMSSM, these two observations can be consistently interpreted as the $\sim 60\gev$ bino-like neutralino DM annihilation~\cite{Abdughani:2021pdc}. 
Nevertheless, the null signal observation from the recent LUX-ZEPLIN (LZ) experiment~\cite{LZ:2022ufs} has probed the majority of parameter space.
Therefore, the remaining parameter space is highly constrained.

\begin{figure}[t]
\centering
\includegraphics[width=0.6\textwidth]{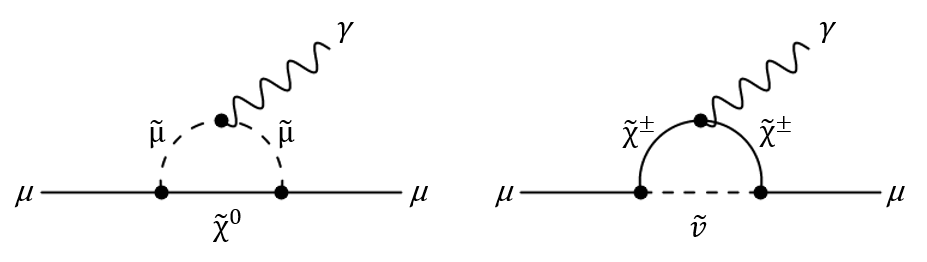}
    \caption{The Feynman diagrams for the one-loop muon $g-2$ corrections with the neutralino-smuon loop and chargino-sneutrino loop.
    }
\label{fig:feynman_g2}
\end{figure}

The muon $g-2$ anomaly reported in Ref.~\cite{2104.03281} can be explained by the contributions of light electroweakinos and sleptons in the SUSY~\cite{Moroi:1995yh,Fowlie:2013oua,Martin:2001st,Abe:2002eq,Stockinger:2006zn,Padley:2015uma,hep-ph/0102122,1610.06587,Abdughani:2019wai,Gu:2021mjd,Baer:2021aax,Aboubrahim:2021ily,Biekotter:2021qbc,Aboubrahim:2021myl,Ali:2021kxa,Wang:2021lwi}.
Since the propagator masses ($m_{\tilde\chi_1^0}$, $m_{\tilde\chi_1^\pm}$, $m_{\tilde\mu}$ and $m_{\tilde\nu_\mu}$) enter the one-loop level of $\delta{a_\mu}$ calculation (as shown in Fig.~\ref{fig:feynman_g2}), 
the geometric mean of the masses of the electroweakinos and sleptons has to be lighter than $\mathcal{O} (400 \gev)$~\cite{Abdughani:2021pdc}. 
Coincidentally, such light electroweakinos and sleptons can also enhance the one-loop contribution of the $W$-boson self-energy~\cite{Domingo:2011uf, 1506.07465}.
Therefore, we re-examine the general behavior of the bino neutralino in the low mass region of $m_{\tilde\chi_1^0}<500$ GeV 
in this work. 
On the other hand, a new singlet $CP$-odd Higgs boson in the NMSSM may play a key role in explaining GCE. 
Although the process of DM annihilation via the SM-like Higgs as a mediator can produce the correct relic density in the early universe, it cannot explain the GCE due to the $p$-wave suppressed annihilation cross section~\cite{1211.1693}. 
Instead, once the singlet $CP$-odd Higgs boson mass approximates two neutralino masses, this singlet-like Higgs resonance annihilation process will be $s$-wave dominated.
Thus, the DM annihilation cross sections at the present time can be around $10^{-26}$ cm$^3 s^{-1}$ required by GCE~\cite{1905.03768}.

The most robust mass limit of the charged SUSY particle is from the LEP. 
Despite the mass spectrum dependency, the LHC can put an even more stringent mass limit on the squark mass, especially for those mass spectra not compressed.  
Based on the latest results of LHC searches for the squark pair production, a squark mass has to be heavier than $1.2\tev$~\cite{Sekmen:2022vzu, ATLAS:2022zwa}. 
All sfermions are involved in the computation of the relic density and $W$-boson mass, but only smuons and sneutrinos contribute to the $\delta{a_\mu}$ calculation in Fig.~\ref{fig:feynman_g2}.
Furthermore, the slepton sector plays a similar role as the squark sector in calculating relic density and $W$-boson mass.    
For simplicity, we decouple the squark sector in this work.

In the framework of MSSM, Ref.~\cite{Yang:2022gvz} has considered CDF II $m_W$, 
muon $g-2$, and DM constraints simultaneously. 
They find that the MSSM can survive all the constraints and agrees with the CDF II $W$-boson mass measurement within $2\sigma$.
Recently, an interesting NMSSM study \cite{Domingo:2022pde} has also performed 
a global analysis of the NMSSM with CDF II $m_W$ measurement. 
Unlike our scenario (bino-like neutralino DM), these authors found that a lighter squark sector is required for a larger $W$-boson mass 
if considering a singlino-like neutralino DM.

In this work, we first perform a comprehensive analysis and then 
test whether the four anomalies or excesses (i.e., CDF II $W$-boson mass, muon $g-2$, GCE, and antiproton) 
may have a common physical origin associated with the bino neutralino in the NMSSM. 
The low-mass bino neutralino can produce correct relic density only with the slepton co-annihilation or $Z/H$-resonance mechanism.
However, the latest LZ experiment~\cite{LZ:2022ufs} gives a rigorous exclusion for DM mass below 100 GeV. 
Finally, we find that only the co-annihilation mechanism can simultaneously explain 
the CDF II W-boson mass excess and g-2 anomaly without violating other experimental constraints. 
Encouragingly, the ongoing LZ, PandaX-4T, and XENON-nT can thoroughly test the surviving parameter space in a few years.

This paper is organized as follows. 
In Sec.~\ref{sec:model}, we briefly introduce the NMSSM model setup and the free parameters chosen in this work. 
The $W$-boson mass prediction in the NMSSM is presented in Sec.~\ref{sec:Wmass},   
and the experimental constraints and measurements are summarized in Sec.~\ref{sec:constraints}. 
In Sec.~\ref{sec:result}, we offer the results and discussion. Finally, we give a conclusion in Sec. ~\ref{sec:conclusion}.

\section{The NMSSM at the electroweak scale}
\label{sec:model}
The superpotential of the scale invariant NMSSM~\cite{Konig:1991tr,Ellwanger:2009dp}, 
which includes a $Z_3$ symmetric gauge singlet chiral superfield $\hat S$, can be written as 
\begin{equation}
    W_\mathrm{NMSSM} = W_\mathrm{MSSM} + \lambda S H_u H_d + \frac{\kappa}{3} S^3,
\end{equation}
where the MSSM component is $W_\mathrm{MSSM}$. 
We denote $\kappa$ and $\lambda$ as dimensionless couplings in the extended Higgs sector. 
The SUSY breaking soft Lagrangian is
\begin{equation}
    -\mathcal{L}_\mathrm{soft} = m^2_{H_u} |H_u|^2 + m^2_{H_d} |H_d|^2 + m^2_s |S|^2 
    + \lambda A_\lambda S H_u H_d + \frac{1}{3} \kappa A_\kappa S^3 + h.c., 
\end{equation}
where $m_{H_u}$, $m_{H_d}$, $m_S$, $A_\lambda$, and $A_\kappa$ are the soft SUSY breaking parameters. 
Adopting the minimization condition for the Higgs potential, 
one can replace $m_{H_u}$, $m_{H_d}$ and $m_S$ with $Z$ boson mass, $\tan\beta$ and $\mu$ parameter. 
Therefore, we have in total six free parameters ($\lambda,~\kappa,~A_\lambda,~A_\kappa,~\mu,~\tan\beta$) for the Higgs sector. 

The DM candidate is the lightest neutralino (i.e., the lightest mass eigenstate of mixed bino $\tilde{B}$, neutral wino $\tilde{W^0}$, neutral up-type higgsino $\tilde{H_u^0}$, neutral down-type higgsino $\tilde{H_d^0}$, and singlino $\tilde{S}$ gauge eigenstates)
\begin{equation}
\tilde\chi^0_{1}=Z_1 \tilde{B} +Z_2 \tilde{W^0}+
Z_3 \tilde{H^0_u} + Z_4 \tilde{H^0_d} +Z_5 \tilde{S}\,.
\label{eq:neutralino}
\end{equation}
Here, $Z_{\rm i}$ for ${\rm i}=(1,~5)$ are coefficients determined by diagonalizing the neutralino mass matrix, 
\begin{eqnarray}
M_{\tilde\chi^0} &=& \left( \begin{array}{ccccc}
   M_1 & 0 & -m_Z c_\beta s_W & m_Z s_\beta s_W  & 0 \\
   &M_2& m_Z c_\beta c_W & -m_Z s_\beta c_W  & 0 \\
	      &   &                  0 & -\mu    & -\lambda v_u\\
	      &  &   &  0 & -\lambda v_d \\
	      &  &   &   & \frac{2\kappa}{\lambda} \mu
	\end{array} \right),
	\label{neutralinomassmatrix}
\end{eqnarray}
where $M_1$ and $M_2$ are the soft bino and wino mass terms, $\mu$ is the effective $\mu$-term. The vacuum expectation values for the Higgs fields $H_u$ and $H_d$ are denoted as $v_u$ and $v_d$, and their ratio is defined as $\tan \beta \equiv v_u /v_d$. Abbreviations $s_\beta$ and $c_\beta$ are $s_\beta=\sin\beta$ and $c_\beta=\cos\beta$, respectively.
Similarly, $s_W$ and $c_W$ are the sine and cosine of weak mixing angle.

\section{$W$-boson mass in the NMSSM}
\label{sec:Wmass}

In this work, we employ the \texttt{FlexibleSUSY-2.7.0}~\cite{1406.2319,1710.03760,2204.05285} to calculate the one-loop corrections to the $W$-boson mass. We briefly summarize the contribution to $m_W$ in the NMSSM. The $W$-boson mass, including both the SM and the BSM contributions, can be expressed as
\begin{equation}
\frac{G_F}{\sqrt{2}}=\frac{\pi\alpha m^2_Z}{2m_W^2(m_Z^2-m_W^2)}(1+\Delta r)\,,
\label{eq:mw}
\end{equation}
where $G_F$ and $\alpha$ are the Fermi constant and fine structure constant, respectively. The $\Delta r$ represents all the non-QED radiative corrections to the muon decay, which depends on the $W$-boson mass. All one-loop diagrams contributing to the $\Delta r$ include the $W$-boson self-energy, vertex, and box corrections. Its form can be written as 
\begin{equation}
\Delta r\simeq\Delta\alpha-\frac{c^2_W}{s^2_W}\Delta\rho\,,
\label{eq:delta_r}
\end{equation}
where the shift of the fine structure constant $\Delta\alpha$ results from the charge renormalization, which involves the contributions from light fermions. 
The $\rho$-parameter describes the neutral and charged weak currents~\cite{Veltman:1977kh}, 
while $\Delta\rho$ is the loop contribution. 
In principle, $\Delta\rho$ can take significant contributions from SUSY particles, but 
$\Delta\alpha$ is a pure SM contribution.
Therefore, additional contributions to $m_W$ is given by 
\begin{equation}
\Delta m_W=\frac{m_W}{2}\frac{c^2_W}{c^2_W-s^2_W}\Delta\rho.
\label{eq:delta_mw}
\end{equation}

\begin{figure}[t]
\centering
\includegraphics[width=0.8\textwidth]{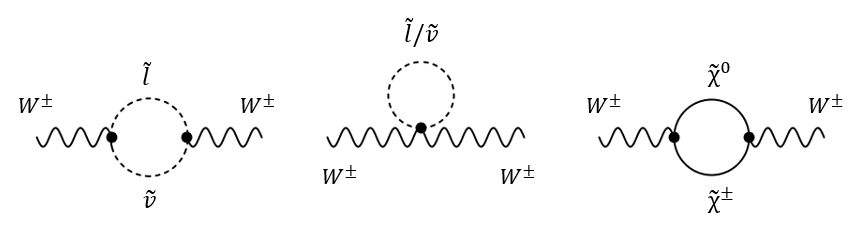}
    \caption{The Feynman diagrams for the one-loop $W$-boson self-energy corrections with the slepton/sneutrino loop and chargino-neutralino loop.}
\label{fig:feynman}
\end{figure}

As seen in Fig.~\ref{fig:feynman}, the dominant contributions at the one-loop in the NMSSM 
result from the sleptons, sneutrinos, and electroweakinos. 
We would like to note that the singlet Higgs can also contribute to the $m_W$ via loop diagrams,  
but a relatively light singlet Higgs mass is required by CDF II W boson measurement~\cite{Domingo:2011uf, 1506.07465}.
However, only a singlet Higgs mass $\mathcal{O}(100\gev)$ can predict the GCE~\cite{Abdughani:2021pdc}. 

\begin{figure}[h]
\includegraphics[width=8.5cm,height=8.5cm]{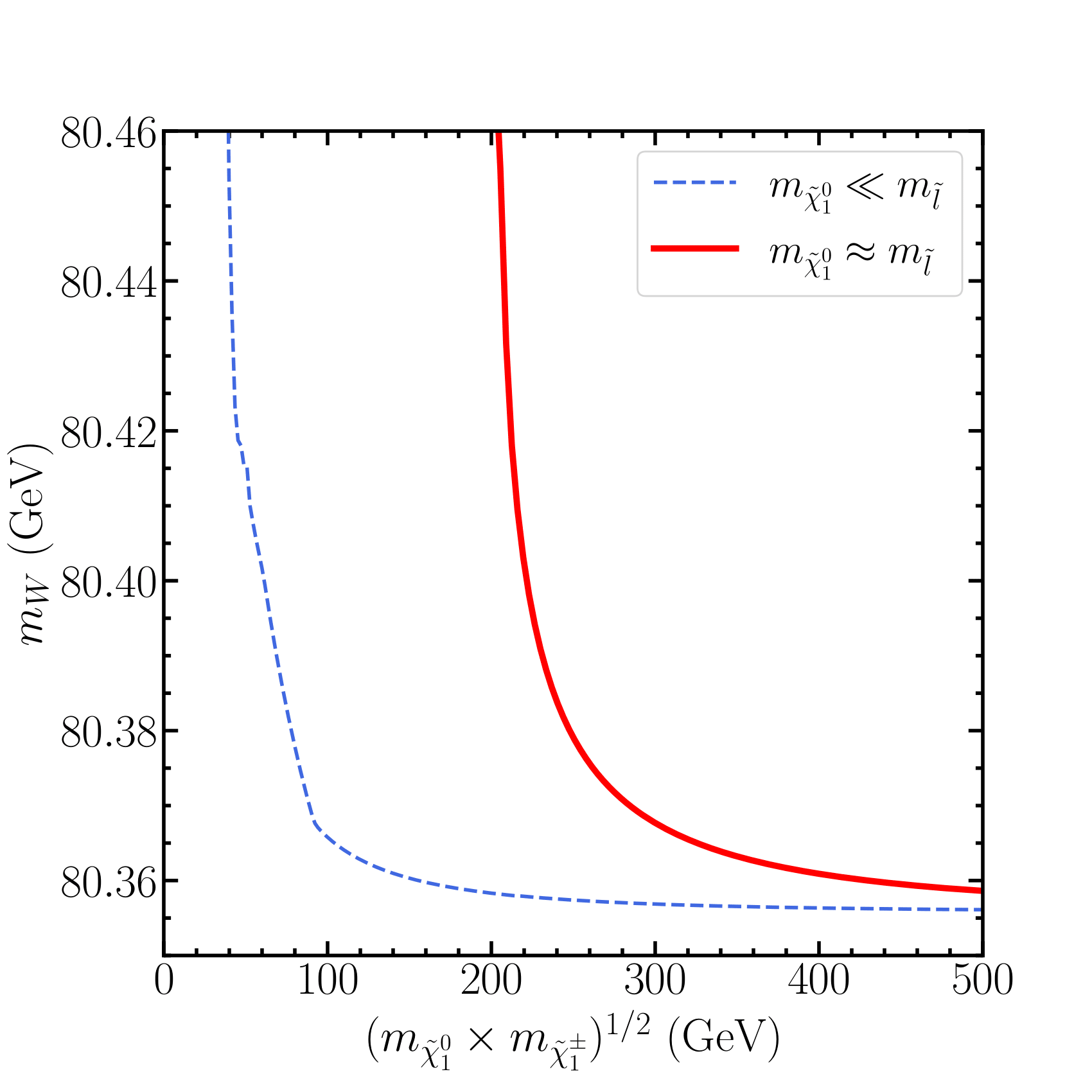}
\caption{The predicted $m_W$ as a function of the electroweakino masses $\sqrt{m_{\tilde \chi^\pm_1}m_{\tilde \chi^0_1}}$. 
The blue dashed line represents only the contribution of chargino-neutralino loop to $m_W$, while the solid red line includes the joint contributions of both chargino-neutralino and slepton loops.
}
\label{fig:testmsusy}
\end{figure}

Let us qualitatively check the loop contributions from the slepton and electroweakino sectors.
As illustrated in Fig.~\ref{fig:feynman}, 
lighter mediator masses ($m_{\tilde{l}}$ and $m_{\tilde{\nu}}$) can enhance the contributions from two left diagrams.  
Similarly, the right diagram is sensitive to the chargino and neutralino masses. 
In Fig.~\ref{fig:testmsusy}, we depict $m_W$ in the function of electroweakino mass 
defined as $\sqrt{m_{\tilde \chi^\pm_1} m_{\tilde\chi^0_1}}$. 
Here, we set the common input parameters $\lambda=3\times 10^{-3}$, $\kappa=10^{-3}$, 
$A_\lambda=460\gev$, $A_\kappa=-0.55\gev$, and $\tan\beta=25$. 
For the blue dashed line, the electroweakino soft mass parameters ($3M_1=M_2=\mu$) are varied from 10 GeV to 500 GeV, 
while other sparticle mass parameters including slepton soft mass $M_{\tilde l_{1,2}}$ are fixed at $3\tev$ for decoupling. 
Oppositely, we vary not only the slepton soft mass parameter but also the electroweakino soft mass parameters 
with a condition $M_{\tilde l_{1,2}}=1.1M_1$ for the solid red line. Meanwhile, all other SUSY parameter settings remain the same as the blue dashed line.

In Fig.~\ref{fig:testmsusy}, only with the contribution from electroweakinos (blue dashed line), 
the required mass to explain CDF measured $m_W$ is $\sim \mathcal{O}(50\gev)$.
However, once the slepton diagrams are included in the $m_W$ correction (solid red line), 
a heavier electroweakino mass $\sim \mathcal{O}(200\gev)$ is needed.
This feature implies that the new contributions from the slepton diagrams are comparable to electroweakino one.  
Hence, one should bear in mind that both $\delta a_\mu$ and $\delta m_W$ can originate from 
the slepton and electroweakino corrections in the NMSSM. 

\section{Experimental constraints and measurements}
\label{sec:constraints}
For the constraints examined in this work, unless specifically mentioned, 
a Gaussian distribution is assumed when the experimental central values ($\xi$), experimental errors ($\sigma$), and theoretical errors ($\tau$) are available. 
Our total chi-square $\chi^2_\mathrm{tot}$ is the sum of $\chi_i^2$, where $i$ runs over 
all the constraints considered in this work, such as SM-like Higgs mass $m_{h_\mathrm{SM}}$, 
B-physics observations including ${\rm BR}(B_s \rightarrow X_s \gamma)$, ${\rm BR}(B^0_s \rightarrow \mu^+ \mu^-)$ 
and ${\rm BR}(B_u \rightarrow \tau \nu)$, DM direct detection, DM relic density $\Omega h^2$, and muon $g-2$. The Gaussian $\chi^2$ is defined as
\begin{equation}
    \chi^2 = \frac{(\theta - \xi)^2}{\sigma^2 +\tau^2}, \label{eq:likelihood}
\end{equation}
where $\theta$ is the calculated theoretical prediction, and
$\xi$ is the experimental central value. 
The uncertainty consists of theoretical error $\tau$ and experimental error $\sigma$.

For most collider constraints (LEP and LHC Higgs searches), 
we apply a $2\sigma$ hard cut likelihood function as implemented in the \texttt{NMSSMTool-5.5.2}. 
On the other hand, several theoretical conditions, such as Landau Pole and tachyonic problem, are also addressed with \texttt{NMSSMTools}.
In what follows, we summarize those experimental constraints/measurements implemented differently or missing in \texttt{NMSSMTools}. 

\begin{itemize}

\item {\textbf{The $W$-boson mass}}\\
    To examine the impact of the new CDF II $m_W$ measurement, we performed two sets of numerical scans of the $W$-boson mass. 
    As a comparison, we denote that the scenario agrees with the SM case as \textbf{PDG2020} 
    by using the SM prediction $m_{W,{\rm SM}}=80.361\pm 0.006\gev$~\cite{ParticleDataGroup:2020ssz}. 
    However, the other scenario \textbf{CDF II} is based on the latest CDF II measurement $m_{W,{\rm CDF}}=80.4335\pm 0.009\gev$~\cite{CDF:2022hxs}.   
    We apply the likelihoods of both scenarios as Gaussian distribution Eq.~\eqref{eq:likelihood}, 
    while the scenario \textbf{CDF II} includes the theoretical uncertainty of $0.006\gev$ given by the SM prediction.

\item {\textbf{Muon magnetic anomaly $\gtwo$}}  \\
    The E989 $\gtwo$ measurement~\cite{2104.03281} has revealed  
    \begin{equation}
       \gtwo =  (2.51 \pm 0.59) \times 10^{-9}. 
    \end{equation}    
    In the one-loop level, the Feynman diagrams in the NMSSM are identical to those in the MSSM, 
    but their relevant couplings can differ. 
    The involving diagrams are $\tilde\chi_1^0$ and $\tilde{\mu}$ exchanging in one loop as well as  
    $\tilde\chi_1^\pm$ and $\tilde{\nu}_\mu$ exchanging in the other loop, as shown in Fig.\ref{fig:feynman_g2}.

\item {\textbf{The charged SUSY particle mass}}\\   
    The most conservative limit of charged SUSY particle mass is 
    from the LEP2~\cite{LEP II}.
   The limit for chargino is $m_{\tilde\chi^\pm}>91.9\gev$, but 
   we require chargino and slepton to be heavier than $100\gev$. 
    Constraints from LHC to low-scale electroweak and sfermion sectors are rather severe.   
    However, the constraints can be weakened when scenarios with compressed mass spectra are considered. 
    We examine our samples against ATLAS $\sqrt{s}=13$ TeV with 139 fb$^{-1}$ luminosity analyses, including 
    two leptons plus missing energy~\cite{1908.08215} and two leptons plus missing energy with jets \cite{1911.12606}.

\item {\textbf{The SM-like Higgs mass and decay}}\\ 
    The latest LHC measured Higgs mass is 
    $m_{h_\mathrm{SM}} = 124.99 \pm 0.18 \pm 0.04\gev$~\cite{ATLAS:2022net}.
    The theoretical uncertainty of Higgs mass is set to be 2 GeV, taking into account the neglected higher-order loop corrections and renormalization scheme differences~\cite{Heinemeyer:2011aa, Fowlie:2012im}.
    The SM Higgs properties can be basically determined by its invisible decay and decay to diphoton. The former constrains the Higgs decay to new particles, 
    particularly DM, while the latter  gives an upper limit to the new charge particles
    involved in the loop calculations. 
    The LHC experimental measurements of the signal strength of Higgs decay to diphoton $R_{\gamma\gamma}$~\cite{ATLAS:2022tnm} 
    and a branching ratio of Higgs decay to invisible particles $R_{\rm inv}$~\cite{ATLAS:2022yvh, ATLAS:2022vkf} are
    \begin{eqnarray}
      R_{\rm inv} &<& 0.145 \pm 0.09 ,      \nonumber \\
      R_{\gamma\gamma} &=& 1.04 \pm 0.10.
    \end{eqnarray} 
    Signal strength of Higgs decay to diphoton is defined as
    \begin{equation}
        R_{\gamma\gamma} \equiv \frac{\sigma_h^{\gamma \gamma}}{\sigma_{h_{\rm SM}}^{\gamma \gamma}} \simeq \frac{{\rm BR}(h \to \gamma \gamma)}{{\rm BR}(h \to \gamma \gamma)_{\rm SM}},
    \end{equation}
    where ${\rm BR}(h \to \gamma \gamma)_{\rm SM} = 2.27 \times 10^{-3}$ \cite{ParticleDataGroup:2020ssz} is the branching ratio of Higgs decay to diphoton final state in the SM.
    
\item {\textbf{B-physics}}\\
    The measurements of ${\rm BR}(B \rightarrow X_s \gamma)$~\cite{HFLAV:2022pwe}, 
    ${\rm BR}(B^0_s \rightarrow \mu^+ \mu^-)$~\cite{LHCb:2021vsc}, and 
    ${\rm BR}(B_u \rightarrow \tau \nu)$~\cite{Zyla:2020zbs} can also constrain SUSY parameter space, i.e.,
    \begin{eqnarray}
           {\rm BR}(B \rightarrow X_s \gamma) &=& (3.49 \pm 0.19 ) \times 10^{-4}, \nonumber\\
           {\rm BR}(B^0_s \rightarrow \mu^+ \mu^-) &=& (3.09 \pm 0.46 \pm 0.15) \times 10^{-9}, \\
           {\rm BR}(B_u \rightarrow \tau \nu) &=& (1.09 \pm 0.24 ) \times 10^{-4}. \nonumber    
           \label{eq:Bphy}
    \end{eqnarray}
    Here, we take $10\%$ uncertainty as theoretical error for ${\rm BR}(B \rightarrow X_s \gamma)$ 
    and ${\rm BR}(B_u \rightarrow \tau \nu)$.

\item {\textbf{DM relic density}}\\
In our analysis, we adopt the Planck 2018 data ($\Omega h^2 = 0.1186 \pm 0.002$)~\cite{Planck:2018vyg} to constrain the parameter space. Moreover, we must consider the uncertainties from the Boltzmann equation solver and the entropy table in the early universe. Hence, we conservatively introduce $\tau=10\%\times \Omega h^2$ as a theoretical error.      
    
\item {\textbf{DM direct detection}}\\
When a neutralino scatters off a nucleon, 
a spin-independent component results from a Higgs boson exchange, while 
a spin-dependent one is through a $Z$-boson exchange. 
In the NMSSM, the $\tilde\chi_1^0$-proton and $\tilde\chi_1^0$-neutron spin-independent couplings are more or less the same, 
but the spin-dependent components are not due to the iso-spin violation. 
The most stringent limit on the $\tilde\chi_1^0$-proton spin-independent cross section $\sigma_{p}^{\rm{SI}}$ is from
the recent LZ collaboration~\cite{LZ:2022ufs}. 
The spin-dependent cross section scatterings off a proton $\sigma_{p}^{\rm{SD}}$ and a neutron $\sigma_{n}^{\rm{SD}}$ are from PICO60~\cite{PICO:2019vsc} and XENON1T~\cite{XENON:2017vdw} experiments, respectively. 
    
\end{itemize}

\section{Numerical Results and Discussions}
\label{sec:result}

To account for the muon $g$-2 and $m_W$ anomalies, 
we need to adjust the electroweakino parameters $M_1$, $M_2$, and slepton mass parameter $M_{\tilde \ell_{1,2}}$.  
In total, we have nine free parameters at electroweak scale, and their ranges are 
\begin{eqnarray}
 10^{-4} < \lambda < 1, &&~10^{-4} < |\kappa| < 2,\nonumber\\
|A_\lambda| < 3\tev, &&~|A_\kappa| < 100\gev,\nonumber\\
30\gev < M_1 < 1\tev, &&~1.5\times M_1 < M_2 < 5\times M_1,\nonumber\\
1.5\times M_1 < |\mu| < 5\times M_1, &&~ 2<\tan\beta<65,\nonumber\\
100\gev < M_{\tilde \ell_{1,2}} < 1\tev .&& 
\label{eq:prior}
\end{eqnarray}   
Other parameters $( M_{\tilde Q}, A_{u,d,b,\ell}, M_3, M_{\tilde \ell_{3}} )$ are set to 3 TeV to decouple,
while $A_t = 4$ TeV to adjust SM-Higgs mass.

We perform the scanning in the ranges summarized above with Markov Chain Monte Carlo (MCMC) method implemented in the \texttt{emcee} \cite{Foreman-Mackey:2012any}. 
The mass spectra and decay information are generated using \texttt{NMSSMTools-5.5.2}~\cite{hep-ph/0508022}.
DM relic density, $\gtwo$, $B$-physics observables, DM-nucleon spin-independent (SI), and spin-dependent (SD) cross sections are obtained using package \texttt{MicrOMEGAs-5.2.11}~\cite{1004.1092}.

In our numerical analysis, the Profiled Likelihood method is adopted.  
Based on all the likelihoods summarized in Sec.~\ref{sec:constraints}, we give two sets of results for $m_{W,\rm{SM}}$ and $m_{W,\rm{CDF}}$, respectively. We collect millions of data points that meet all constraints. 
For a two-dimension plot, the 95\% confidence ($2\sigma$) region is defined by $\delta\chi^2<5.99$ under the assumption of approximate Gaussian likelihood.
The global minimum $\chi^2$ for the \textbf{CDF II} (\textbf{PDG2020}) scenario is $8.89$ ($8.65$). 
Such similar results indicate 
that there is no preference for \textbf{PDG2020} or \textbf{CDF II} scenarios in the NMSSM.  
Thus, the NMSSM can reasonably reproduce both data sets without conflicting with other experimental constraints.

\begin{figure}
\includegraphics[width=8.1cm,height=8.1cm]{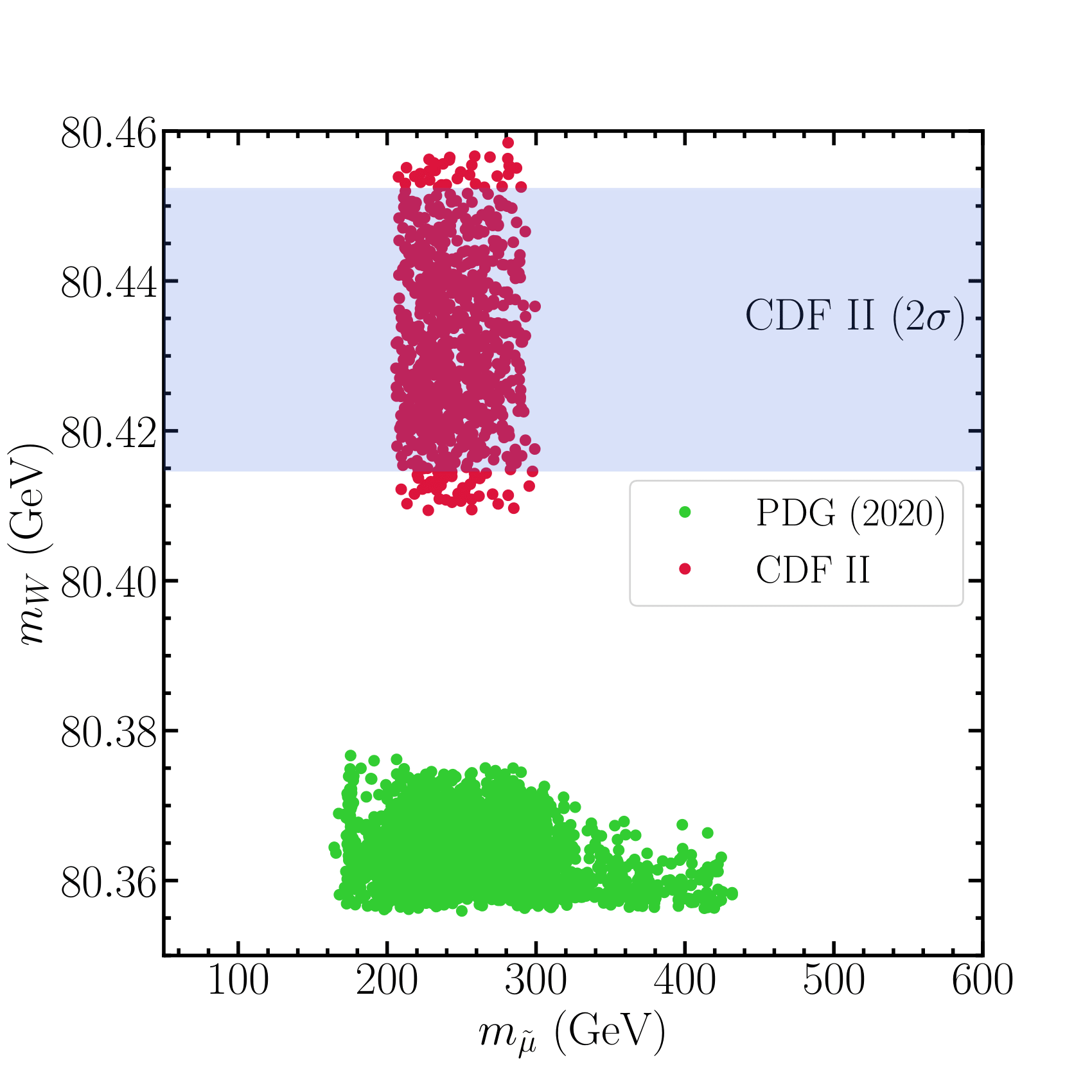}
\includegraphics[width=8.1cm,height=8.1cm]{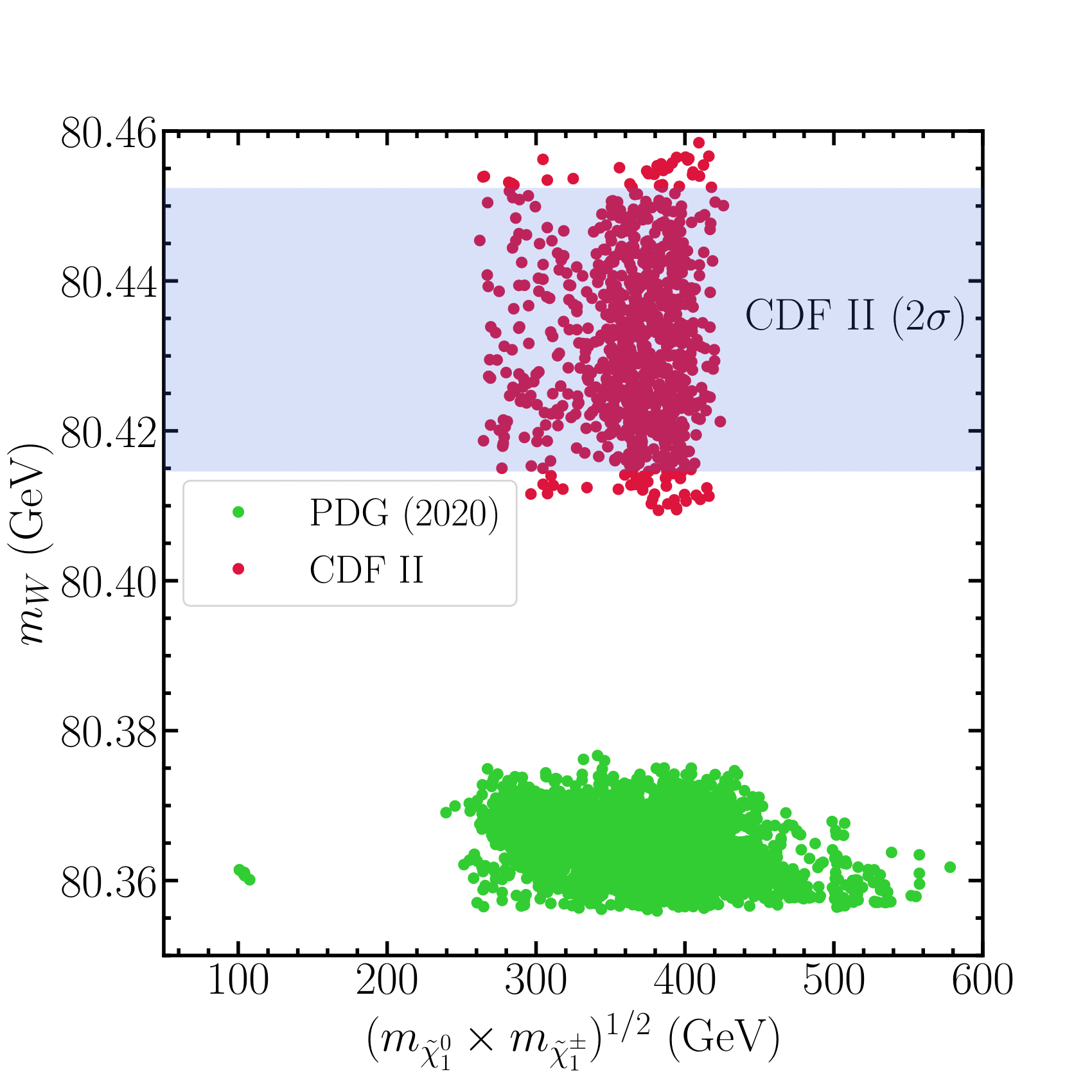}
\caption{The $2\sigma$ likelihood distributions with the CDF II $m_W$ (red points) and SM $m_W$ (green points) constraint, respectively. 
The left and right panels show the expected value of $m_W$ as a function of $m_{\tilde\mu}$ and 
$\sqrt{m_{\tilde \chi^\pm_1} m_{\tilde\chi^0_1}}$, respectively. 
The light blue band is for the $2\sigma$ region of the CDF II $m_W$ measurement.
}
\label{fig:mw}
\end{figure}

In Fig.~\ref{fig:mw}, we show the predicted value of $m_W$ for $m_{\tilde\mu}$ (left panel) and $(m_{\tilde \chi^\pm_1} \times m_{\tilde\chi^0_1})^{1/2}$ (right panel). 
The group in red represents the \textbf{CDF II} scenario, while the group in green represents SM $m_W$ labeled \textbf{PDG2020}.
We can draw two general conclusions from Fig.~\ref{fig:mw}. First, both slepton and electroweakino loops contribute to the $W$-boson mass.
Second, explaining the CDF II $W$ mass requires lighter sparticles, such as the $m_{\tilde\mu}$ clusters in the range of $\sim 200-300$ GeV.

In SUSY, the main contributions to the muon $g-2$ come from the neutralino-smuon and chargino-sneutrino loops, as shown in Fig.~\ref{fig:feynman_g2}.
The expressions of one-loop corrections to $a_\mu$ can be seen in Ref.~\cite{Martin:2001st}.
Assuming all sparticles have a universal SUSY mass scale $M_{\rm{SUSY}}$, the SUSY contribution to muon $g-2$ can be expressed as~\cite{Moroi:1995yh}
\begin{equation}
\delta a_\mu^{\rm{SUSY}}=14\times 10^{-10} \times \tan\beta\left(\frac{100\rm{GeV}}{M_{\rm{SUSY}}}\right)^2. 
\label{eq:g2}
\end{equation}    
We can see that a large $\tan\beta$ enhances the SUSY contributions.
On the other hand, the lightest neutralino $\tilde\chi^0_1$ (LSP) can interact with nuclei via the exchange of Higgs bosons in the DM direct detection. The corresponding coupling of the LSP with the Higgs boson can be approximated as 
\begin{equation}
C_{H\tilde\chi^0_1\tilde\chi^0_1} \simeq -\sqrt{2}g_1 Z^2_{1}\frac{M_Zs_W}{\mu}\frac{M_1/\mu+\sin{2\beta}}{1-(M_1/\mu)^2},
\label{eq:CHxx}
\end{equation}    
where $Z_{1}$ is the bino component of $\tilde\chi^0_1$ mass eigenstate. 
The DM-proton spin-independent scattering cross section $\sigsip$ depends on the ratio $M_1/\mu$, and the value of $\sin 2\beta$ which is inversely proportional to $\tan\beta$. 
In other words, $\sigsip$ can be enhanced by enlarging $M_1/\mu$ and reducing $\tan\beta$.

\begin{figure} 
\includegraphics[width=8.1cm,height=8.1cm]{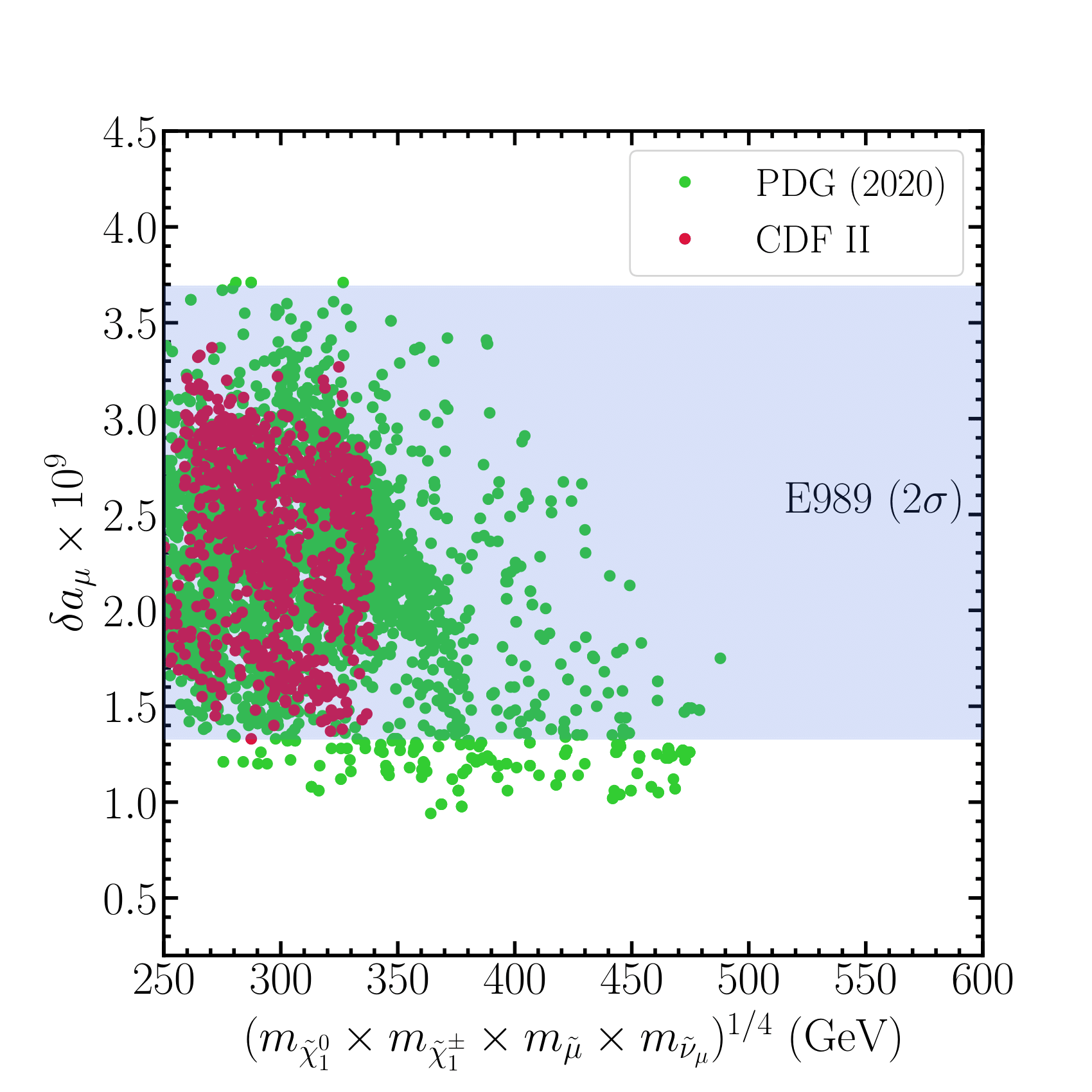}
\includegraphics[width=8.1cm,height=8.1cm]{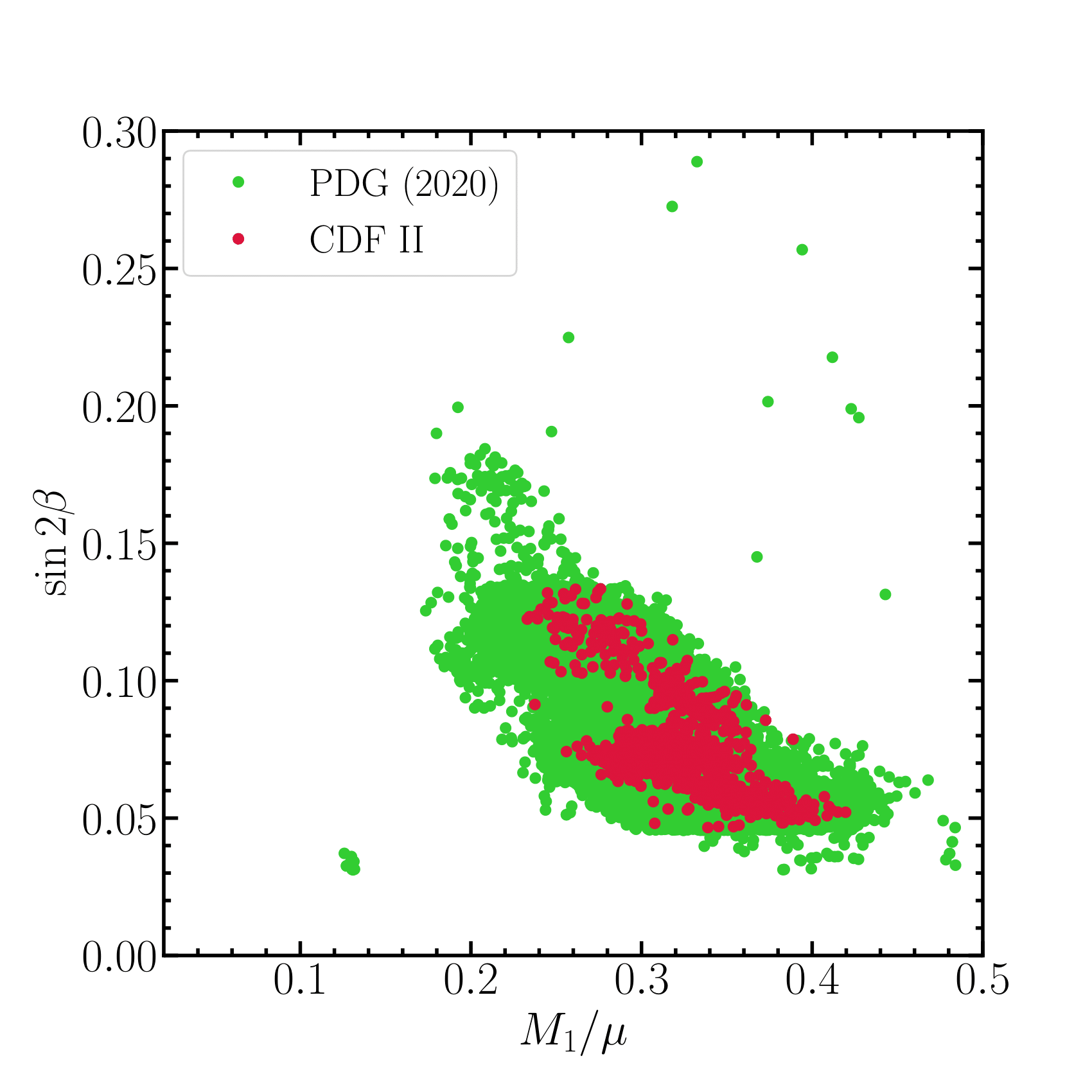}
\includegraphics[width=8.1cm,height=8.1cm]{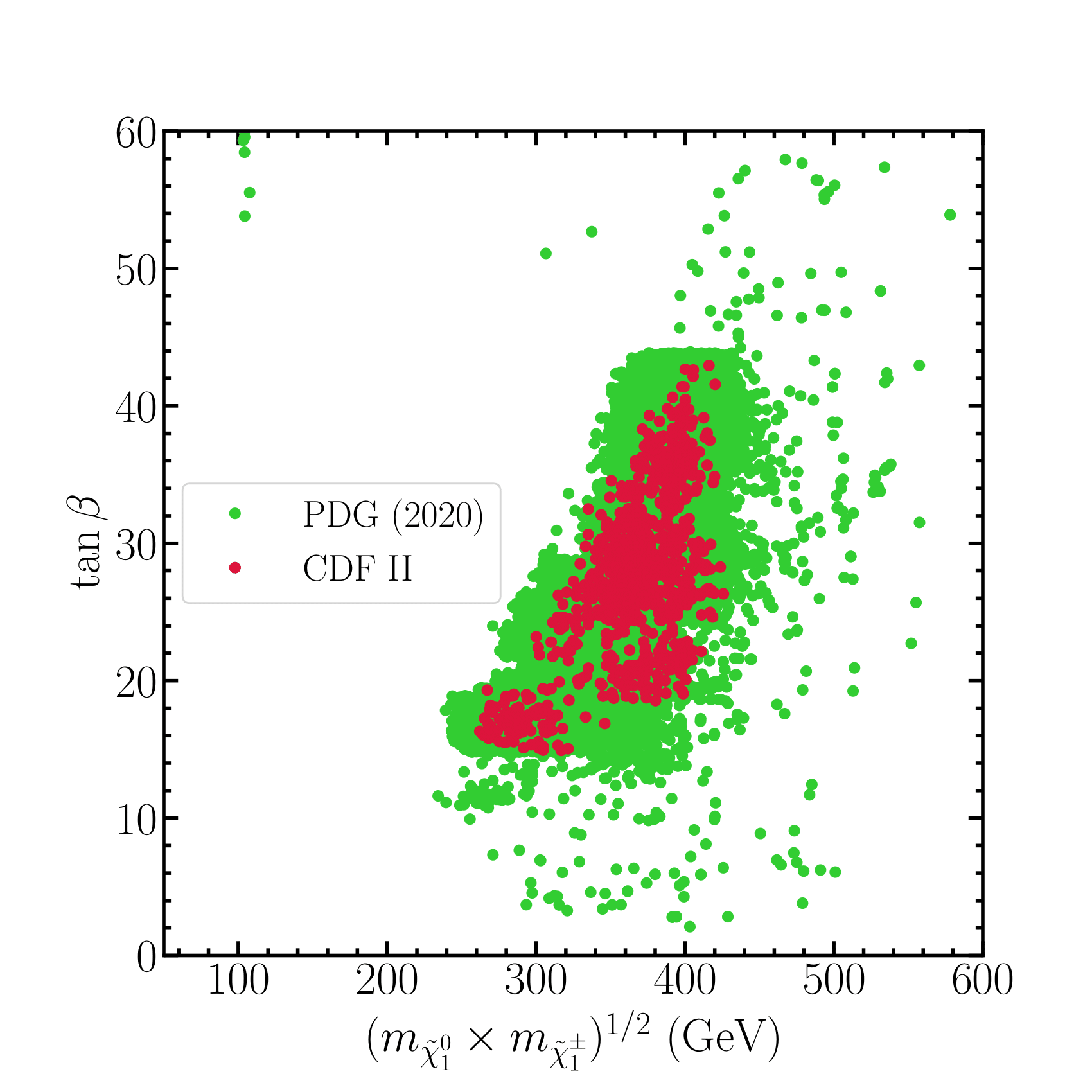}
\includegraphics[width=8.1cm,height=8.1cm]{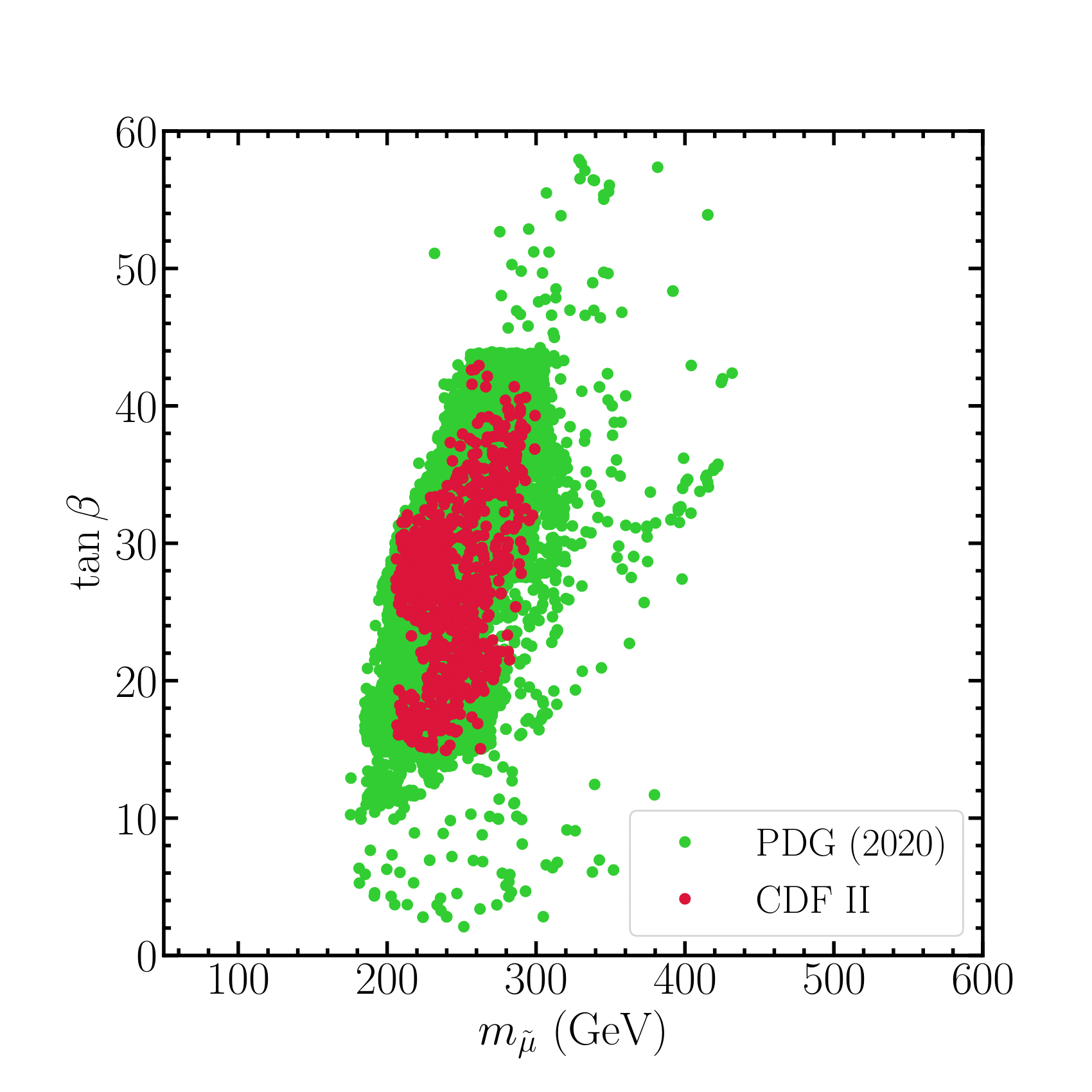}
\caption{The $2\sigma$ allowed regions for PDG2020 (green points) and CDF II (red points) scenarios.
The upper left panel presents the predicted value of $\delta a_\mu$ for the geometric mean of the masses of $m_{\tilde\chi^{0}_{1}}$, $m_{\tilde\chi^{\pm}_{1}}$, $m_{\tilde{\mu}}$, and $m_{\tilde{\nu}_{\mu}}$, 
where the light blue region is the E989 $2\sigma$ allowed region.
The upper right panel displays the $\sin2\beta$ as the function of $M_1/\mu$.
The lower left and right panels show the expected value of $\tan\beta$ versus $\sqrt{m_{\tilde \chi^\pm_1} m_{\tilde\chi^0_1}}$ and $m_{\tilde\mu}$, respectively.}
\label{fig:g2}
\end{figure}

In the upper left panel of Fig.~\ref{fig:g2}, we show 
the predicted value of $\delta a_\mu$ for 
the geometric mean of the masses of $m_{\tilde\chi^{0}_{1}}$, $m_{\tilde\chi^{\pm}_{1}}$, $m_{\tilde{\mu}}$, 
and $m_{\tilde{\nu}_{\mu}}$. 
We confirm the conclusion in \cite{Abdughani:2021pdc} that the geometric mean of the masses of the electroweakino and slepton shall be less than $\mathcal{O}(500\gev)$
to reproduce the $\delta a_\mu$ measurement. 
Since the electroweakino and slepton are also involved in the $W$-boson mass correction, 
as illustrated in Fig.~\ref{fig:feynman}, 
the CDF II $m_W$ measurement plays a crucial role in setting a mass upper limit at $\mathcal{O}(340\gev)$.

One can see from Eq.~\eqref{eq:g2} that the measured value $\delta a_\mu$ of E989  
requires a larger value of $\tan\beta$. 
However, to explain CDF $m_W$ measurement, an even larger value $\tan\beta>15$ is needed,  
as demonstrated in two lower planes of Fig.~\ref{fig:g2}. 
This interesting result can be understood as follows. 
Both $M_1$ and $\mu$ control the electroweakino one-loop correction of $m_W$. 
For larger values of $M_1$ and $\mu$, the heavier mediator masses, $m_{\tilde\chi^0}$ and $m_{\tilde\chi^\pm}$, 
can suppress the corresponding one-loop correction of $m_W$. 
Hence, a larger $\tan\beta$ is required to enhance 
the ${W^\pm\tilde\chi^0\tilde\chi^\pm}$ coupling $C_{W^\pm\tilde\chi^0\tilde\chi^\pm}$~\cite{Jungman:1995df}.

\begin{figure}
\includegraphics[width=8.1cm,height=8.1cm]{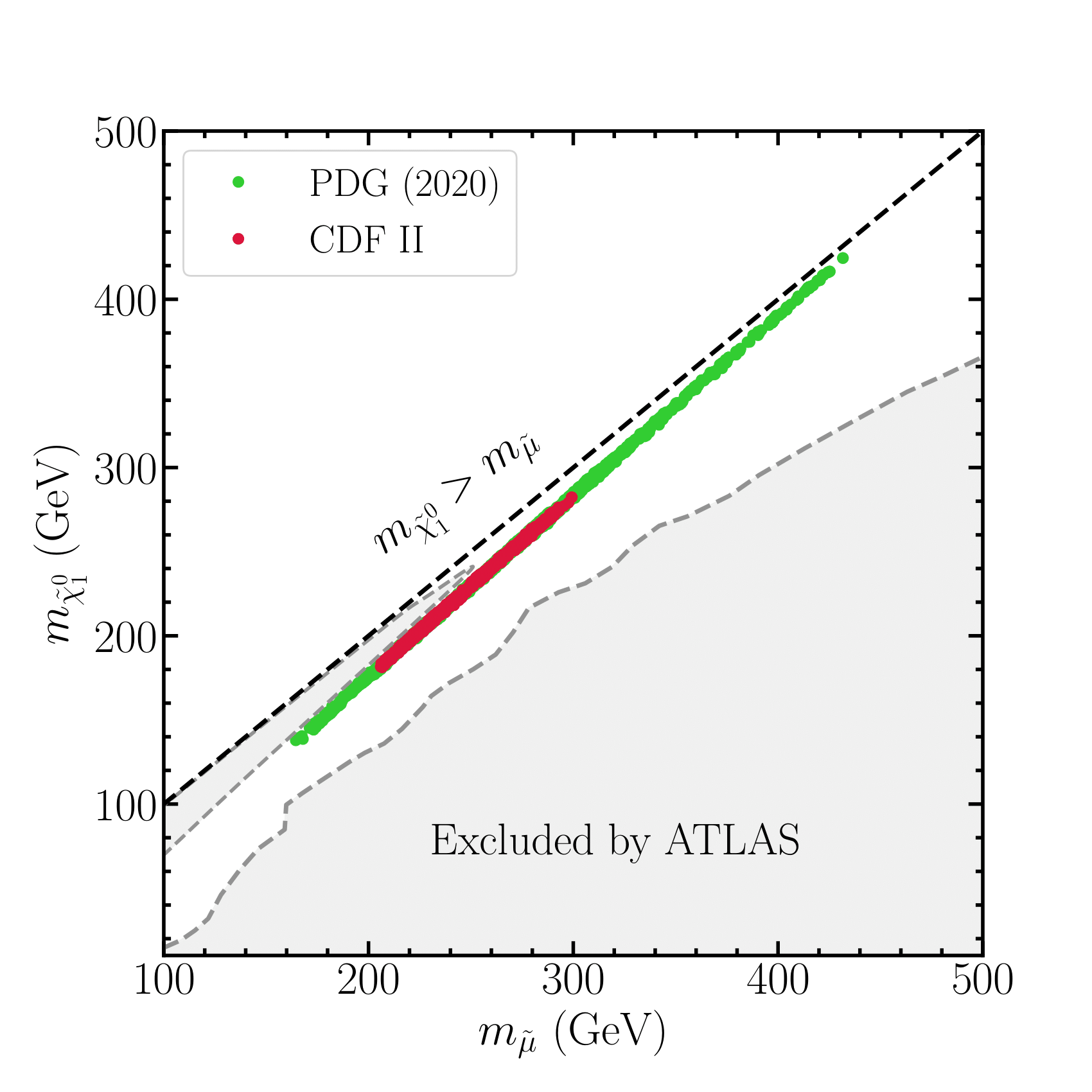}
\includegraphics[width=8.1cm,height=8.1cm]{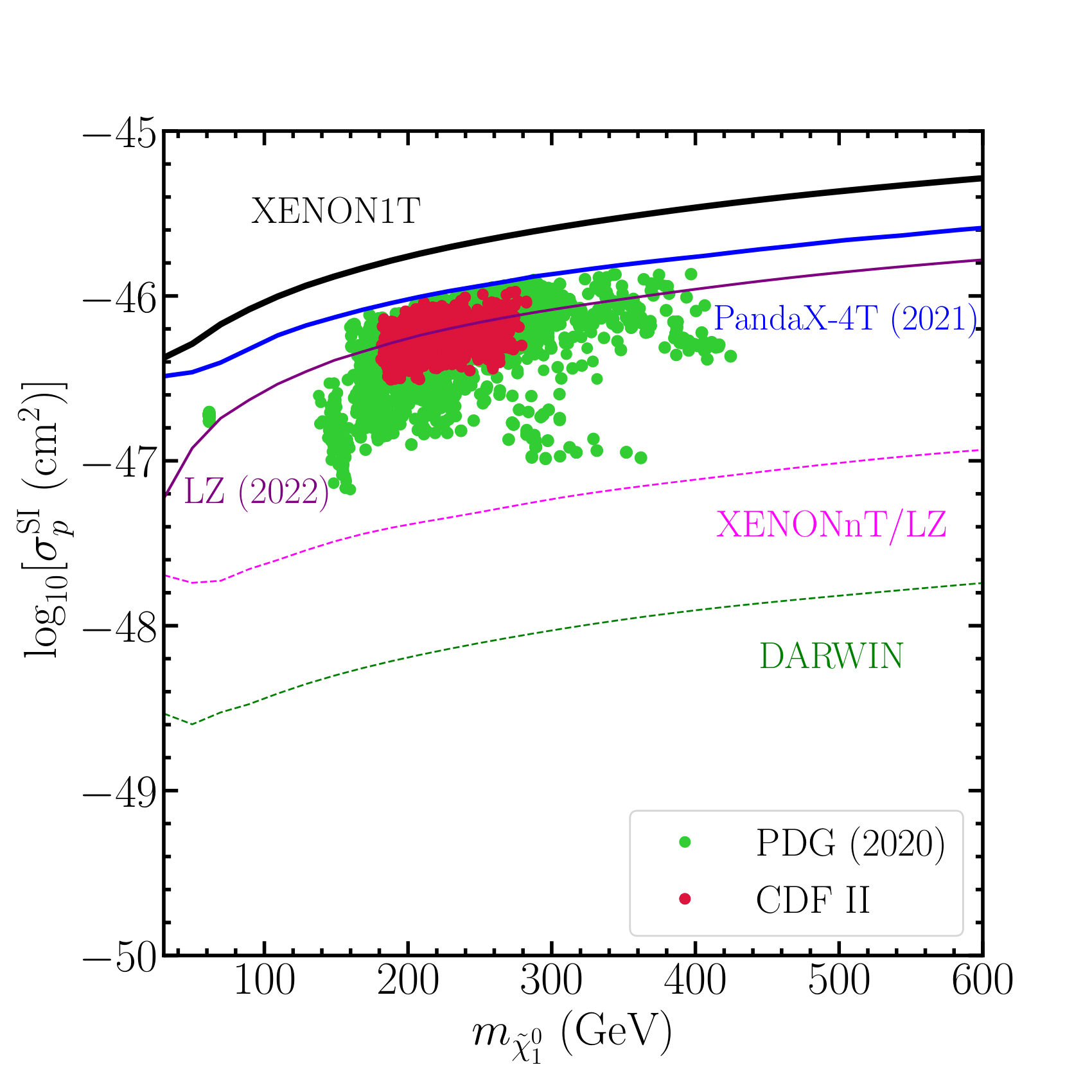}
\caption{The left panel displays the allowed $m_{\tilde{\mu}}$ as a function of the dark matter mass $m_{\tilde\chi_{1}^{0}}$. The ATLAS sets tight bounds, and the surviving points are in the co-annihilation region.
The right panel shows the predicted value of $\sigsip$ for the dark matter mass $m_{\tilde\chi_{1}^{0}}$.
The three solid lines represent the current constraints set by XENON1T (solid black line)~\cite{XENON:2017vdw}, PandaX-4T (sold blue line)~\cite{PandaX-4T:2021bab}, and LZ (sold purple line)~\cite{LZ:2022ufs}, while the two dashed lines are the projected sensitivities of XENONnT/LZ (dashed pink line)~\cite{XENON:2020kmp,1509.02910} and DARWIN (dashed green line)~\cite{Schumann:2015cpa}.}
\label{fig:exp}
\end{figure}

In the upper right panel of Fig.~\ref{fig:g2}, we show the correlation between $M_1/\mu$ and $\sin 2\beta$ 
as given in Eq.~\eqref{eq:CHxx}. 
In the range $10<\tan\beta<60$, the $2\sigma$ allowed region corresponds to $\sin 2\beta < M_1/\mu$. 
Together with Eq.~\eqref{eq:CHxx}, we find that the DM direct detection limits enforce a small value of $M_1/\mu$.   
Interestingly, this implies a promising prospect of DM direct detection 
(see the right panel of Fig.~\ref{fig:exp}).

We also consider the exclusion from the null results of searching for the sparticles with two leptons plus missing energy at the 13 TeV LHC with the integrated luminosity of 139 fb$^{-1}$ ~\cite{1908.08215,1911.12606}. We examine our samples, which include three main signal processes: (i) $pp \to \tilde l^+_{L,R}\tilde l^-_{L,R}$, (ii) $pp\to \tilde\chi_1^+ \tilde\chi_1^-$, and (iii) $pp\to \tilde\chi^\pm_1 \tilde\chi_2^0 $.
In the left panel of Fig.\ref{fig:exp}, we show the favored regions of $m_{\tilde\chi_{1}^{0}}$ versus $m_{\tilde{\mu}}$, where the light gray regions are the tight bounds of the ATLAS.
Clearly, there is a tight correlation between $m_{\tilde\chi_{1}^{0}}$ and $m_{\tilde{\mu}}$, which suggests a co-annihilation origin.
We would like to point out that the latest LZ experiment~\cite{LZ:2022ufs} has excluded the $Z/H$-resonance region after considering CDF II $W$-boson mass (see the right panel of Fig.\ref{fig:exp}).
Strikingly, the parameter space favored by the current data can be entirely probed by the ongoing/upcoming DM direct detection experiments such as the PandaX-4T~\cite{PandaX:2018wtu}, XENONnT/LZ~\cite{XENON:2020kmp,1509.02910}, and DARWIN~\cite{Schumann:2015cpa}. 
The corresponding bino-like neutralino DM mass is  $180\gev < m_{\tilde\chi_{1}^{0}}< 280\gev$.

\begin{figure}
\includegraphics[width=8.1cm,height=8.1cm]{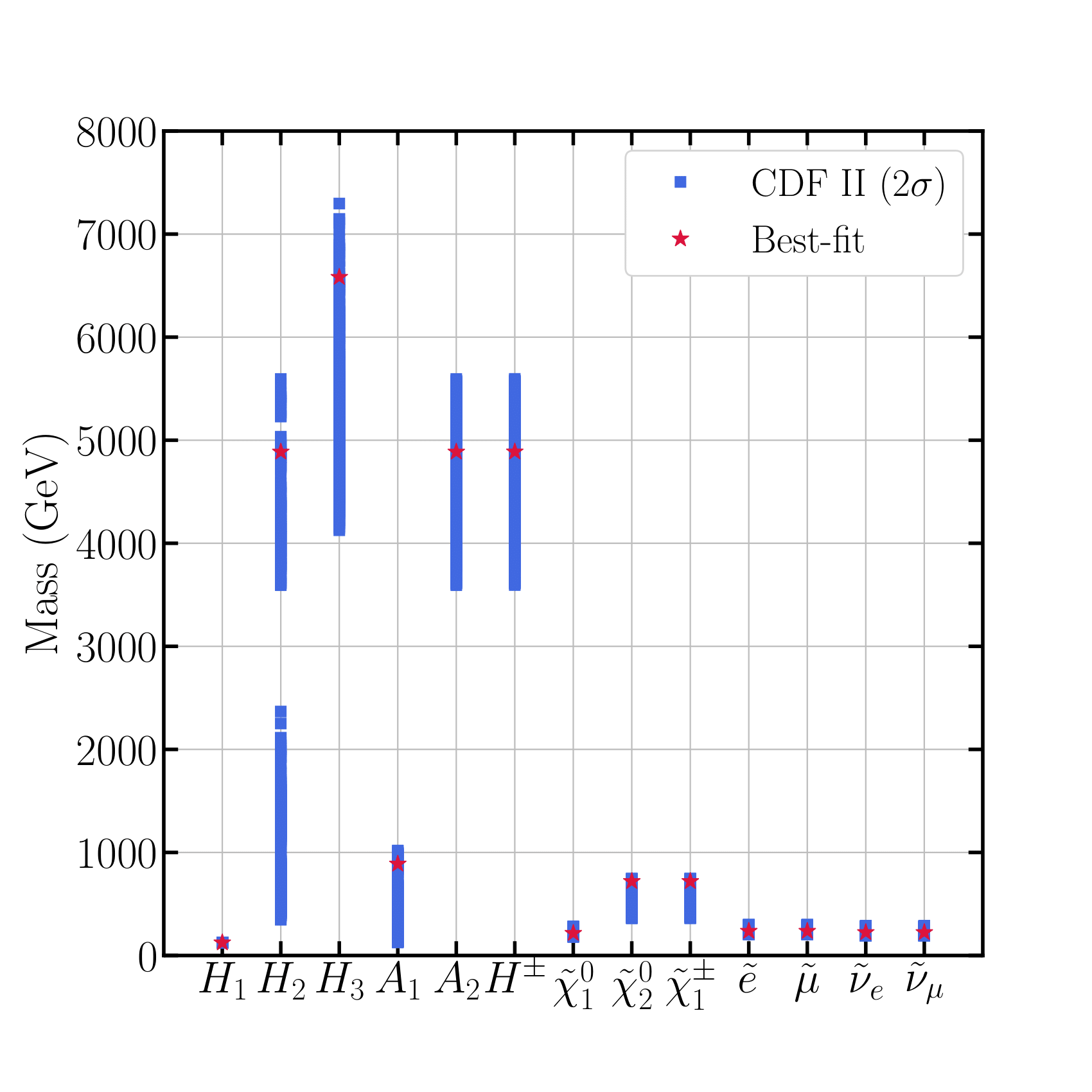}
\includegraphics[width=8.1cm,height=8.1cm]{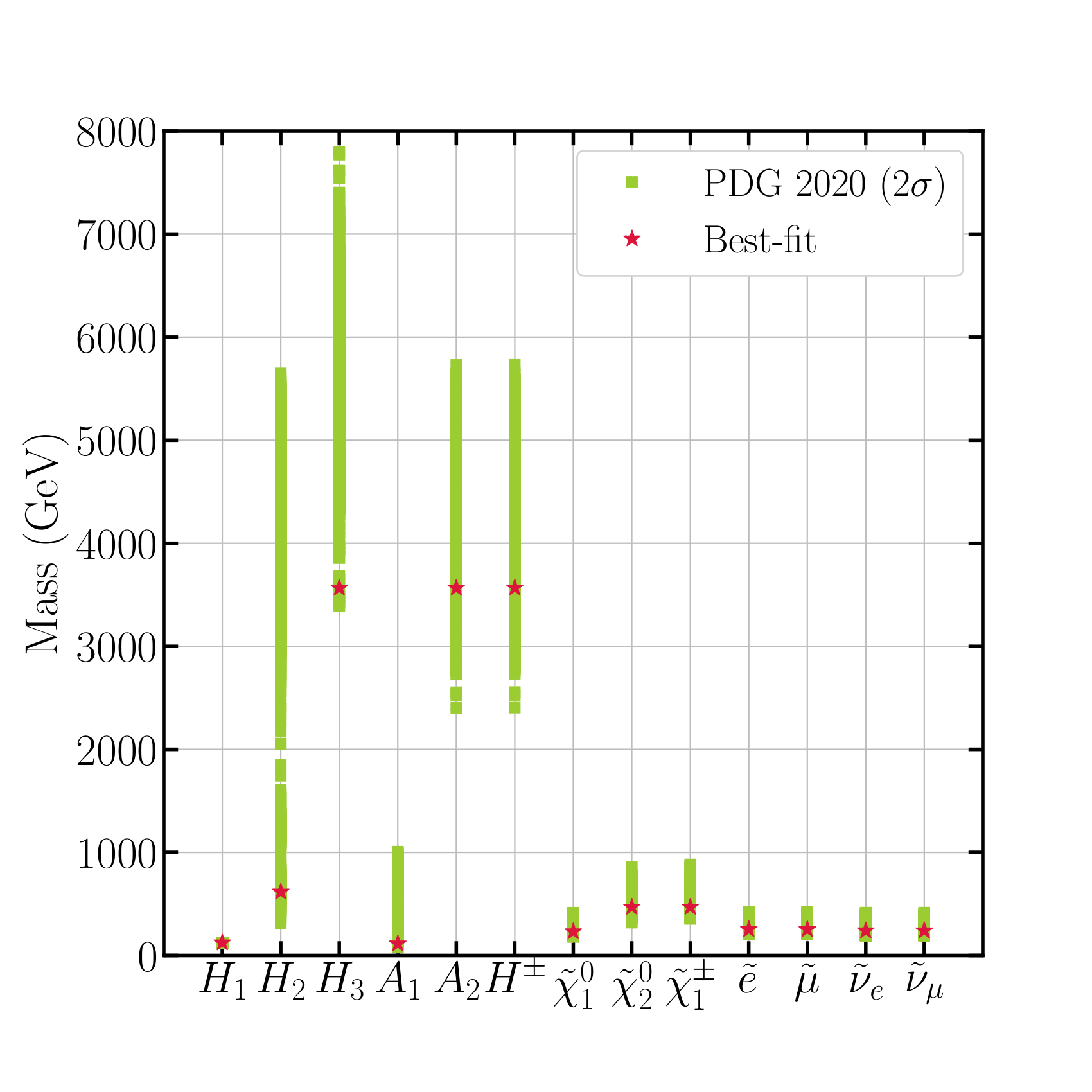}
\caption{Best-fit mass spectra of the NMSSM in two likelihood combinations (red stars) and $2\sigma$ (blue bars for CDF II $m_W$ and green bars for SM $m_W$) confidence intervals. Squarks with a physical mass of about 3 TeV are not shown here, which have been set to decouple in our calculation. }
\label{fig:mass}
\end{figure}

The particle mass spectra for the best-fit points and $2\sigma$ confidence regions are shown in Fig~\ref{fig:mass}. 
The left panel corresponds to the \textbf{CDF II} scenario, while 
the right panel is for the \textbf{PDG2020} scenario.

\section{Conclusion}
\label{sec:conclusion}

We would like to summarize our findings.
Based on the bino-like DM scenario, the CDF II $m_W$ and muon $g-2$ anomalies can be simultaneously explained by the contributions from the light electroweakinos and sleptons in the NMSSM.
In this work, we compute one-loop correction to the $m_W$ via \texttt{FlexibleSUSY-2.7.0}~\cite{1406.2319,1710.03760,2204.05285}~\footnote{Recently, the newly released NMSSMTools-6.0.0~\cite{Ellwanger:2004xm, Ellwanger:2005dv, nmssmtools6.0.0} can also be used to compute $m_W$. We have checked our results with this package and verified our conclusion.}.
Motivated by the favored region to explain the recent $a_{\mu}$ anomaly, 
we only focus on the lighter electroweakino and slepton sectors without mass splitting between the left and right sparticles.
We successfully find the solution to explain both muon $g-2$ and CDF $ W$ mass anomalies without conflicting with other experimental constraints.
The corresponding bino-like neutralino DM mass is in the range of $\sim 180-280$ GeV.
We should emphasize that the solution expected to satisfy GCE by a singlet-Higgs resonance has probably been ruled out since the latest LZ experiment~\cite{LZ:2022ufs} shows a powerful ability to exclude DM with mass below 100 GeV.
Although $Z/H$-resonance regions have been excluded, the slepton co-annihilation is still viable.
The DM annihilation cross sections at the present time are less than $10^{-27}$ cm$^3 s^{-1}$ due to the co-annihilation suppression, which is challenging for DM indirect detection experiments.
However, the favored DM mass region can soon be entirely probed by ongoing DM direct detection experiments like PandaX-4T, XENONnT, LZ, and DARWIN.

\section*{Acknowledgments}
We appreciate
Peter Athron and Ulrich Ellwanger for their insightful suggestions and helpful discussions. This work was supported in part by the National Natural Science Foundation of China (No. 11921003 and No. U1738210),  by China Postdoctoral Science Foundation (2020M681757), and by the Key Research Program of the 
Chinese Academy of Sciences (No. XDPB15).

\end{document}